\documentclass[conference]{IEEEtran}
\IEEEoverridecommandlockouts
\usepackage{amsmath,amssymb,amsfonts}
\usepackage{graphicx}
\usepackage{algpseudocode}
\usepackage{array}
\usepackage{makecell}
\usepackage{textcomp}
\usepackage[caption=false,font=footnotesize]{subfig}
\usepackage{booktabs}
\usepackage{threeparttable}
\usepackage{tabularx}
\usepackage{stfloats}
\usepackage{placeins}
\usepackage{multirow}
\usepackage{soul}
\usepackage{xcolor}
\usepackage{arydshln}
\usepackage{verbatim}
\usepackage{adjustbox}
\usepackage{cite}
\setulcolor{red}
\definecolor{titleblue}{RGB}{0,0,255}
\definecolor{skyblue}{RGB}{135,206,235}

\newcolumntype{C}[1]{>{\centering\arraybackslash}p{#1}}

\def\BibTeX{{\rm B\kern-.05em{\sc i\kern-.025em b}\kern-.08em
T\kern-.1667em\lower.7ex\hbox{E}\kern-.125emX}}
\renewcommand{\IEEEkeywordsname}{Keywords}

\begin{document}

\title{A Novel Byte-Level Flow-to-Image Encoding Method for Network Intrusion Detection Systems\\
}
\author{
Ziyu Mu*, Zihui Yan, Xiyu Shi, and Safak Dogan\\
\textit{Institute for Digital Technologies}\\
\textit{Loughborough University London, London E20 3BS, UK}\\

\{z.mu, z.yan, x.shi, s.dogan\}@lboro.ac.uk
}

\maketitle

\begin{abstract}
Network-based Intrusion Detection Systems (IDS) are predominantly trained on tabular flow records, whose one-dimensional representations limit convolutional architectures from exploiting inter-feature spatial correlations. This paper presents a novel byte-level flow-to-image encoding method that converts each network-flow record into a fixed-size RGB image. Continuous features are serialised using IEEE-754 single-precision format and packed sequentially into pixels along an inverted-L shaped trajectory, while discrete features are mapped to byte values and placed contiguously in the middle image row's centre. The encoding is deterministic and reversible, preserving a fixed spatial layout across all samples. Four IDS models are evaluated on NSL-KDD and UNSW-NB15 datasets with both flow and image-based configurations. The image-based representation yields consistent accuracy gains of up to 15.6\% and 12.8\% for binary and multi-classification on UNSW-NB15, and up to 3.5\% and 3.2\% on NSL-KDD, highlighting the potential of byte-level visual encoding to strengthen AI-driven intrusion detection in local computer networks.\\

\end{abstract}

\begin{IEEEkeywords}
\textit{Intrusion Detection System (IDS), Flow-to-Image Encoding, Network Traffic Security}
\end{IEEEkeywords}

\section{Introduction}
The security of local computer networks has become increasingly critical as cyber threats grow in sophistication and frequency. Network-based Intrusion Detection Systems (IDS) serve as a frontline defence mechanism, relying on Machine Learning (ML) and Deep Learning (DL) models to classify network traffic as malicious or benign in real time~\cite{mehavilla2026evaluating}. The existing methods train detection models such as Support Vector Machines (SVM), and neural network architectures directly on one-dimensional (1D) tabular feature vectors~\cite{11192261, 11391661}. While effective for known attack patterns, 1D models treat features independently or through sequential operations, which limit their capacity to capture local cross-feature dependencies that may carry discriminative information.

The success of Convolutional Neural Networks (CNN) in image classification has prompted several studies to convert network traffic into two-dimensional (2D) image representations~\cite{11095369, 11032121}. However, existing encoding methods typically rely on dimensionality reduction or heuristic spatial arrangements that do not preserve original feature values, potentially discarding fine-grained numerical information that is critical for reliable intrusion detection.

This paper presents a byte-level flow-to-image encoding method that addresses these issues through a deterministic and reversible conversion. Each flow record is mapped to a fixed-size 32×32 RGB image: discrete features are encoded as integer byte indices and logged into assigned RGB channels, while continuous features are serialised using IEEE-754~\cite{8766229} single-precision format into four bytes and packed sequentially into pixels along an inverted-L shaped trajectory. The unused positions are zero-padded. This mapping ensures that the same feature always occupies the same spatial region, enabling convolutional filters to learn consistent local patterns.

The main contributions of this work are as follows:

\begin{itemize}
\item A novel byte-level flow-to-image encoding method is proposed that deterministically and reversibly converts tabular flow records into 32×32 RGB images, preserving the original feature values without loss of numerical precision.
\item Experimental evaluation on NSL-KDD~\cite{5356528} and UNSW-NB15~\cite{7348942} with their original training and testing splits demonstrates that the proposed encoding consistently improves detection accuracy across both shallow and deep IDS model architectures.
\item The proposed method provides a practical and deployable approach to strengthening AI-driven intrusion detection in local computer networks, requiring no modification to existing model architectures.

\end{itemize}

The rest of the paper is structured as follows. Section II presents the related work. Section III described the proposed encoding method and IDS models. Section IV outlines the experimental settings and datasets. Section V provides the experimental evaluation and discussion of the results. Section VI concludes the paper with directions for future work.

\section{Related Work}

\subsection{Deep Learning-based Intrusion Detection}

Securing local computer networks against evolving intrusion threats has driven significant research interest in AI-enabled detection approaches. Xue et al.~\cite{xue2025hae} introduced HAE-HRL, combining a CNN-Gated Recurrent Unit (GRU) autoencoder for feature selection with an enhanced ResNet-Long Short-Term Memory (LSTM) classifier, achieving binary classification accuracy of 95.7\%, 94.9\%, and 96.7\% on NSL-KDD, UNSW-NB15, and CICIDS2018, respectively. Bensaoud and Kalita~\cite{BENSAOUD2025103770} proposed a framework integrating Self-Organising Maps, Deep Belief Networks, and Autoencoders optimised via Particle Swarm Optimisation, reporting accuracy up to 99.99\% on NSL-KDD and UNSW-NB15, though under specific experimental conditions that may not generalise across evaluation protocols. Farrukh et al.~\cite{FARRUKH2024103982} presented AIS-NIDS, a packet-level system combining closed- and open-set classifiers with incremental learning to adapt to unknown attack classes, though performance exhibited degradation as the number of classes increased.

Besides standard classification models, generative approaches have been explored to mitigate data scarcity in network security. A Boundary Equilibrium Generative Adversarial Network (BEGAN)-based system~\cite{park2022enhanced} generated synthetic minority-class samples and evaluated several ML and DL models on NSL-KDD, with reported accuracy ranging from 72.1\% to 82.6\%. A Wasserstein Generative Adversarial Network with Gradient Penalty (WGAN-GP)-based approach~\cite{mu2024information} yielded accuracy improvements of up to 2.3\% on NSL-KDD through data augmentation, though the underlying 1D tabular input structure remained unchanged.

These approaches demonstrate competitive detection performance, yet the majority operate on 1D feature representations, limiting the capacity of convolutional architectures to exploit cross-feature dependencies.

\subsection{Flow-to-Image Encoding Methods for IDS}

As network traffic volumes in local computer networks continue to grow, several studies have explored converting traffic data into image representations to leverage the pattern recognition capabilities of 2D convolutional architectures. Farrukh et al.~\cite{Farrukh2023SENetI} proposed SeNet-I, serialising consecutive packets into three-channel RGB images, achieving F1-scores of 96\% and 83\% for flow- and packet-based multi-classifications. However, reliance on packet payload limits applicability to flow-level datasets. El-Ghamry et al.~\cite{el2023optimized} arranged selected features via recursive feature elimination from NSL-KDD into 9×9×3 colour images, with VGG16 achieving 97.89\% binary accuracy. However, image dimensions are constrained by the number of selected features, and the spatial arrangement carries no inherent semantic meaning. Lilhore et al.~\cite{lilhore2024cognitive} mapped flow records to 24-bit RGB images for a MobileNetV3-SVM hybrid, achieving over 98\% precision on CICIDS2017, CICIDS2018, and UNSW-NB15. Sun and Wang~\cite{SUN2025113236} proposed ICNN-ID, which converts network traffic records into grid-structured images using a fixed-random feature mapping with cyclic feature replication and applies modified LeNet networks for classification, reaching 89.97\% multi-classification accuracy on NSL-KDD. This result benefits from a forced minority inclusion strategy that guarantees all rare attack samples are retained in the training set \cite{11045551}, which improves minority class coverage during training. 

A common limitation across all these methods is the absence of a principled byte-level encoding scheme. Most rely on normalisation to the 0--255 pixel range~\cite{ghadermazi2024towards, 9862964} or heuristic spatial arrangements~\cite{yu2025feature, 9916253, SUN2025113236}, introducing quantisation errors or arbitrary feature placements that discard fine-grained numerical information. Furthermore, few methods rely on dataset-specific preprocessing strategies to achieve competitive performance, which limits their generalisability across datasets with different class distributions. In contrast, our proposed method serialises continuous features using IEEE-754 single-precision format and maps discrete features to byte values, producing deterministic 32×32 RGB images in a fully reversible manner. This design preserves the numerical fidelity of traffic feature representations, making it particularly suited to network security monitoring scenarios where fine-grained feature information is critical for reliable threat detection in local computer networks.

\section{Proposed Approach}

\begin{figure*}[t]
\centering
\includegraphics[width=0.75\textwidth]{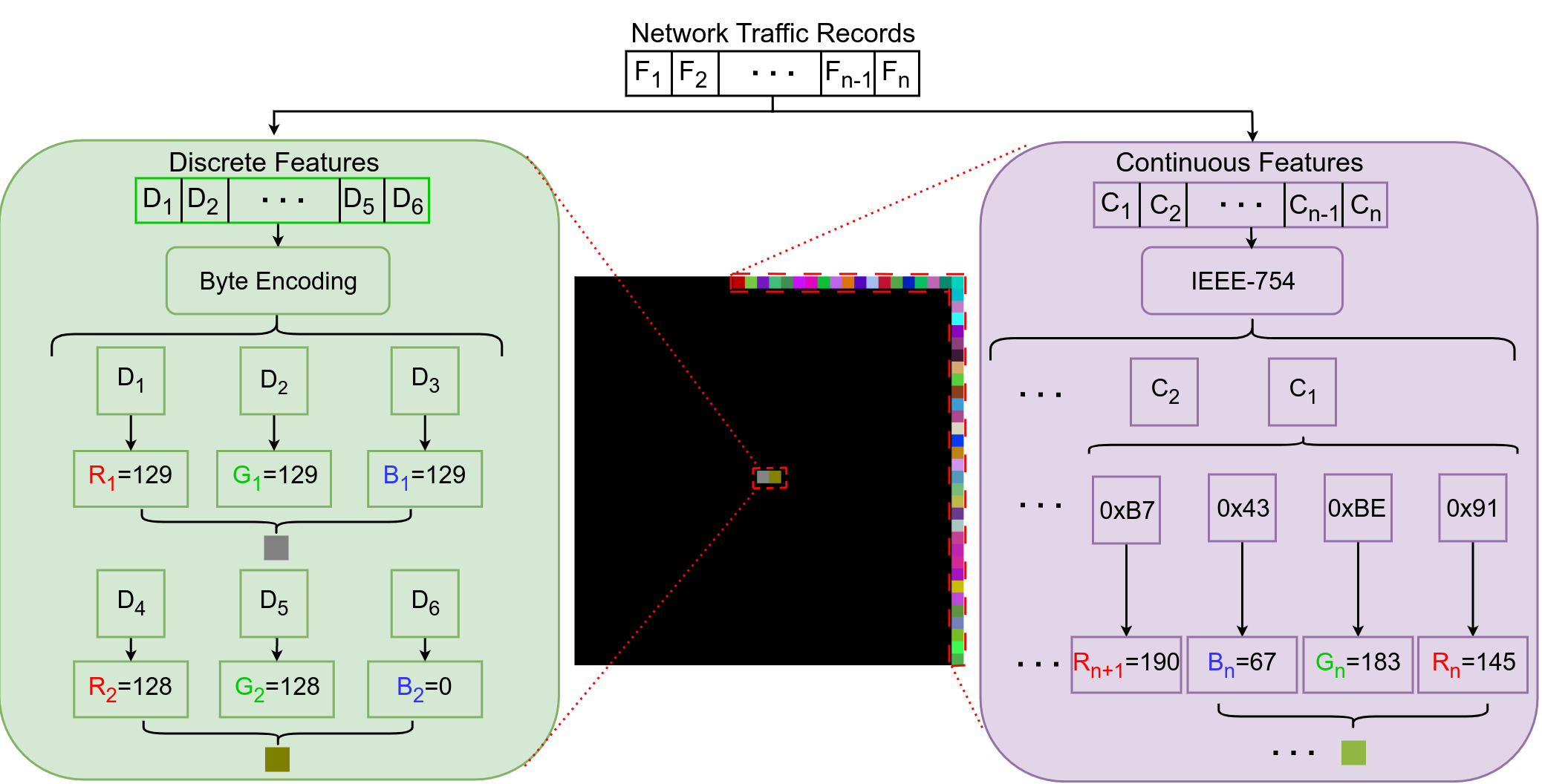}
\caption{Illustration of the proposed flow-to-image encoding process in the UNSW-NB15 dataset. Continuous features are serialised and mapped along an inverted-L shaped trajectory. Since a pixel holds 3 bytes and a feature has 4 bytes, a single feature's bytes are distributed across the RGB channels of adjacent pixels, encouraging convolutional filters to capture both intra-feature byte patterns and inter-feature spatial correlations. Discrete features are encoded as integer byte indices and written contiguously onto the horizontally centred region of the middle row, providing a fixed spatial anchor that is spatially separated from the continuous feature region.}
\label{fig:l_shape_encoding}
\end{figure*}

\subsection{Flow-to-Image Encoding Method}
The primary motivation of the proposed encoding is to transform tabular network traffic records into a spatial representation that is more compatible with the local pattern extraction mechanism of convolutional models. Training IDS models directly on raw feature vectors presents two practical limitations: first, local cross-feature dependencies are weakly captured in a 1D feature space; second, one-hot encoding~\cite{9512057} of categorical fields introduces high-dimensional sparsity and poor alignment with continuous feature representations \cite{11369258}. To address these limitations, a deterministic flow-to-image encoding mapping is designed with three objectives: locality preservation for continuous features, modality-aware spatial placement for categorical versus continuous information, and fixed-size encoding independent of sample-specific feature distributions.

Specifically, each network flow record is mapped to a fixed-size RGB image through byte-level serialisation and predefined spatial layout rules. This renders the encoding reproducible and model-agnostic, while enabling convolutional receptive fields to capture local byte-level interaction patterns from neighbouring pixels. The same pipeline is applied to both UNSW-NB15 and NSL-KDD using their original training and testing splits, producing one encoded image per flow record.

\paragraph{Feature Partition and Statistical Preparation}
The partitioning of features into continuous and categorical types is dataset-specific. UNSW-NB15 comprises 37 continuous features and 5 categorical features, while NSL-KDD comprises 38 continuous features and 3 categorical features. Let $x_{i,c}$ denote feature $c$ of sample $i$.
Continuous features are standardised using statistics estimated exclusively from the training split:
\begin{equation}
z_{i,c}=\frac{x_{i,c}-\mu_c}{\sigma_c+\epsilon},
\end{equation}
where $z_{i,c}$ denotes the standardised value of continuous feature $c$ for sample $i$, $\mu_c$ and $\sigma_c$ are the mean and standard deviation of feature $c$ computed from the training split, and $\epsilon=10^{-8}$ is a small constant for numerical stability. Missing continuous values are imputed with $\mu_c$, yielding $z_{i,c}=0$ after normalisation. For categorical features, a vocabulary is constructed from the training set with an explicit unknown-category (UNK) token assigned index 0, and each category is mapped to an integer index rather than one-hot encoding. This design avoids the high-dimensional sparsity introduced by one-hot encoding and keeps categorical and continuous information mutually compatible under byte-level serialisation.

\paragraph{Byte Encoding and Spatial Layout}
All images use a black background and a fixed size $S \times S$, where $S=32$. The continuous feature vector is first cast to \texttt{float32} and serialised in little-endian byte order, then logged along an inverted-L trajectory: starting from $(S\!-\!1,S\!-\!1)$ upward to $(0,S\!-\!1)$, followed by leftward filling along the top row to $(0,0)$, as illustrated in Fig.~\ref{fig:l_shape_encoding}.

Importantly, since each pixel holds three bytes (R, G, B) while each continuous feature comprises four bytes, the boundaries of features and pixels are intentionally misaligned. Consequently, a single feature's bytes are distributed across the RGB channels of adjacent pixels. This encourages convolutional filters to capture both the byte-level structure within individual features and spatial relationships between adjacent features, as the kernels are required to aggregate information dispersed across both channel and spatial dimensions. Each RGB pixel contributes 3 bytes, and the total byte capacity of the inverted-L path is $(2S-1)\times 3$. For $S=32$, this capacity is 189 bytes, which safely covers both the 148-byte UNSW-NB15 continuous payload ($37\times 4$) and 152-byte NSL-KDD continuous payload ($38\times 4$).

The spatial layout is consistent with the design objectives stated above. Logging continuous features along a deterministic inverted-L path preserves byte adjacency from serialisation to image space, enabling nearby convolutional kernels to capture short-range cross-feature interactions. Placing categorical indices on a fixed central row provides a stable spatial anchor and prevents interleaving of discrete index patterns with floating-point byte patterns. This yields explicit modality separation at shallow layers and controlled modality fusion at deeper layers as receptive fields expand.

\begin{figure*}[t]
    \centering
    \includegraphics[width=0.11\textwidth]{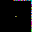}\hspace{4pt}
    \includegraphics[width=0.11\textwidth]{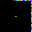}\hspace{4pt}
    \includegraphics[width=0.11\textwidth]{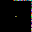}\hspace{4pt}
    \includegraphics[width=0.11\textwidth]{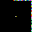}\hspace{4pt}
    \includegraphics[width=0.11\textwidth]{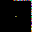}
    \vspace{1pt}\\
    \makebox[0.11\textwidth]{\small Normal}\hspace{4pt}
    \makebox[0.11\textwidth]{\small DoS}\hspace{4pt}
    \makebox[0.11\textwidth]{\small Reconnaissance}\hspace{4pt}
    \makebox[0.11\textwidth]{\small Shellcode}\hspace{4pt}
    \makebox[0.11\textwidth]{\small Worms}\\
    \vspace{1pt}
    \centerline{\small (a) UNSW-NB15}
    \vspace{0.4em}
    \includegraphics[width=0.11\textwidth]{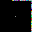}\hspace{4pt}
    \includegraphics[width=0.11\textwidth]{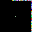}\hspace{4pt}
    \includegraphics[width=0.11\textwidth]{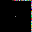}\hspace{4pt}
    \includegraphics[width=0.11\textwidth]{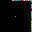}\hspace{4pt}
    \includegraphics[width=0.11\textwidth]{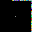}
    \vspace{1pt}\\
    \makebox[0.11\textwidth]{\small Normal}\hspace{4pt}
    \makebox[0.11\textwidth]{\small DoS}\hspace{4pt}
    \makebox[0.11\textwidth]{\small Probe}\hspace{4pt}
    \makebox[0.11\textwidth]{\small R2L}\hspace{4pt}
    \makebox[0.11\textwidth]{\small U2R}\\
    \vspace{1pt}
    \centerline{\small (b) NSL-KDD}
    \caption{Representative 32$\times$32 encoded images generated by the proposed byte-level flow-to-image encoding, showing one sample per category from UNSW-NB15 (top row) and NSL-KDD (bottom row). Continuous features are encoded along the inverted-L path, discrete features occupy the centre row, and unused positions are zero-padded (black).}
    \label{fig:encoding_examples}
\end{figure*}

Categorical indices are encoded as 1-byte values automatically promoted to 2 bytes if the vocabulary size exceeds 255, and logged in the centre row at position ($\lfloor S/2 \rfloor$). The categorical byte stream is horizontally centred and contiguously recorded in RGB channel order. Representative encoded samples from both datasets are shown in Fig.~\ref{fig:encoding_examples}.

\subsection{IDS Models}

Four IDS models are evaluated with both 1D (flow) and 2D (image) configurations, sharing the same architectural design, number of layers, filter sizes, and hidden dimensions across modalities. The primary differences lie in the dimensionality of convolutional and pooling operations and reshaping of feature maps for recurrent components.

\begin{itemize}

\item \textbf{CNN}~\cite{park2022enhanced}: two convolutional layers with 64 filters, each followed by max pooling and batch normalisation, with dropout 0.1 after the first block and a 16-unit fully connected classifier.

\item \textbf{CNN-LSTM}~\cite{bamber2025hybrid}: the same convolutional extractor as CNN, with feature maps reshaped into a sequence processed by a two-layer LSTM; the final time step is passed through layer normalisation before classification.

\item \textbf{CNN-BiLSTM}~\cite{jouhari2024lightweight}: a single convolutional layer with 64 filters, max pooling, and batch normalisation, followed by a bidirectional LSTM (hidden size of 16) with mean pooling, dropout 0.2, and a 64-unit fully connected layer.

\item \textbf{ResNeXt-50}~\cite{ZHAO2026104783}: ResNeXt-50 32$\times$4d configuration with layers [3,4,6,3], 32 groups, and base width of 4. The 2D version uses a CIFAR-style TorchVision stem; the 1D version adapts all convolutions to one dimension with the same bottleneck configuration.
\end{itemize}

\section{Experiments}

\subsection{Dataset}

Two widely used benchmark datasets are adopted for evaluation, namely NSL-KDD and UNSW-NB15. Both datasets provide original training and testing splits, which are used without modification in all experiments to avoid split contamination and ensure reproducibility.

The NSL-KDD dataset contains 41 features comprising 3 nominal and 38 numeric types, of which 3 are treated as categorical and 38 as continuous in the encoding. The binary classification task distinguishes Normal from Attack, while the multi-classification task includes five classes: Normal, Remote to Local Attack (R2L), Probe, Denial of Service Attack (DoS), and User to Root Attack (U2R).

The UNSW-NB15 dataset contains 49 features. Following standard preprocessing, 7 non-traffic fields are excluded, yielding 42 features used in the encoding. For the binary classification task, all attack categories are merged into a single Attack class. For the multi-classification task, five classes are selected following~\cite{10312801}: Normal, DoS, Reconnaissance, Shellcode, and Worms. The flow-based preprocessing pipeline and baseline experimental results follow the methodology established in~\cite{mu2026gmasawgangpnoveldatagenerative}.

Table~\ref{tab:datasets_labels} lists the data distribution for both datasets with their original splits.
\begin{table}[t]
  \centering
  \begin{threeparttable}
    \caption{Dataset Statistics}
    \label{tab:datasets_labels}
    \scriptsize
    \renewcommand{\arraystretch}{1.05}
    \setlength{\tabcolsep}{10pt}
    \begin{tabular}{l l c c}
      \toprule
      & & \multicolumn{2}{c}{Number of samples} \\
      \cmidrule(lr){3-4}
      Dataset & Class 
      & \makecell{Training set} 
      & \makecell{Test set} \\
      \midrule
      \multirow{5}{*}{NSL-KDD}
        & Normal          & 67343 & 9711 \\
        & R2L             & 995   & 2885 \\
        & Probe           & 11656 & 2421 \\
        & DoS             & 45927 & 7460 \\
        & U2R             & 52    & 67 \\
      \midrule
      \multirow{5}{*}{UNSW-NB15}
        & Normal          & 56000 & 37000 \\
        & DoS             & 12264 & 4089 \\
        & Reconnaissance  & 10491 & 3496 \\
        & Shellcode       & 1133  & 378 \\
        & Worms           & 130   & 44 \\
      \bottomrule
    \end{tabular}
  \end{threeparttable}
\end{table}

\subsection{Experiment Setup}

A unified training set is applied across all IDS models and modalities. The Adam optimiser is used with a learning rate of $1 \times 10^{-4}$ and a batch size of 64. The CNN-, CNN-LSTM-, and CNN-BiLSTM-based IDS models are trained for 100 epochs, while ResNeXt-50 is trained for 30 epochs due to its faster convergence. Since neither dataset provides an original validation set, 10\% of the training split is held out via stratified sampling for validation purposes. For the flow-based modality, features are scaled to $[-1.0, 1.0]$ using min-max normalisation fitted on the training split. For the image-based modality, continuous features are standardised via z-score normalisation using training-set statistics, and geometric augmentation is not applied in order to preserve the spatial positions of encoded features. Each experiment is repeated ten times with seeds 42 to 51, and results are reported as mean $\pm$ standard deviation. All IDS models are implemented in PyTorch 1.10.2 with CUDA v11.3 and cuDNN v8.1, trained on a Linux workstation with Ubuntu 20.04.5 LTS, an NVIDIA GeForce RTX 3090 Ti GPU, and an Intel Core i7-12700KF CPU.

\section{Results and Discussions}

\subsection{Evaluation Metrics}

In this work, two metrics are used to evaluate the performance of the four IDS models: Accuracy and F1-score. Here, TP, TN, FP, and FN denote the number of True Positives, True Negatives, False Positives, and False Negatives, respectively. Accuracy and F1-score are defined as follows:
\begin{equation}
\text{Accuracy} = \frac{\text{TP} + \text{TN}}{\text{TP} + \text{FP} + \text{FN} + \text{TN}}
\label{formula.accuracy}
\end{equation}

\begin{equation}
\text{F1-score} = \frac{2 \times \text{TP}}{2 \times \text{TP} + \text{FP} + \text{FN}}
\label{formula.f1}
\end{equation}

\subsection{Binary Classification}
\begin{table*}[htbp]
    \centering
    \begin{threeparttable}
    \caption{Binary Classification Results on UNSW-NB15 and NSL-KDD with Different IDS models}
    \label{tab2}
    \scriptsize
    \begin{tabular*}{\textwidth}{@{\extracolsep{\fill}} l l l
        C{1.0cm} C{1.0cm}
        C{0.8cm} C{1.0cm}
        C{0.8cm} C{1.0cm}}
    \toprule
    \textbf{Dataset} & \textbf{Training Data} & \textbf{IDS Model} 
    & \multicolumn{2}{c}{\textbf{Overall}} 
    & \multicolumn{2}{c}{\textbf{Normal}} 
    & \multicolumn{2}{c}{\textbf{Attack}} \\
    \cmidrule(lr){4-5} \cmidrule(lr){6-7} \cmidrule(lr){8-9}
    & & 
    & \textbf{Acc.(\%)} & \textbf{F1(\%)} 
    & \textbf{Acc.(\%)} & \textbf{F1(\%)} 
    & \textbf{Acc.(\%)} & \textbf{F1(\%)} \\
    \midrule
     
    \multirow{9}{*}{\textbf{UNSW-NB15}} 
        & \multirow{4}{*}{Original} 
        & CNN \cite{park2022enhanced}  & 81.8$\pm$4.7 & 68.8$\pm$4.4 & 89.1$\pm$6.5 & 88.8$\pm$3.3 & \textbf{48.1}$\pm$8.5 & 48.7$\pm$6.2 \\
        & & CNN-BiLSTM \cite{jouhari2024lightweight} & 80.7$\pm$0.4 & 61.9$\pm$0.4 & \textbf{91.9}$\pm$0.6 & 88.7$\pm$0.3 & 29.2$\pm$1.2 & 35.0$\pm$0.8 \\
        & & CNN-LSTM \cite{bamber2025hybrid} & 81.0$\pm$0.9 & 62.8$\pm$1.9 & \textbf{91.9}$\pm$0.8 & 88.6$\pm$0.6 & 31.0$\pm$3.5 & 36.7$\pm$3.4 \\
        & & ResNeXt-50 \cite{ZHAO2026104783} & \textbf{82.0}$\pm$1.8 & 66.2$\pm$3.5 & 91.4$\pm$1.3 & 89.3$\pm$1.1 & 38.5$\pm$6.4 & 43.2$\pm$6.0 \\
    \cmidrule(lr){2-9}
        & \multirow{4}{*}{Image} 
        & CNN \cite{park2022enhanced} & 96.6$\pm$0.1 & 94.5$\pm$0.2 & 96.6$\pm$0.1 & 97.9$\pm$0.1 & 96.9$\pm$0.3 & 91.1$\pm$0.3 \\
        & & CNN-BiLSTM \cite{jouhari2024lightweight} & 96.3$\pm$0.1 & 94.0$\pm$0.2 & \textbf{96.7}$\pm$0.1 & 98.0$\pm$0.0 & 97.1$\pm$0.1 & 91.4$\pm$0.2 \\
        & & CNN-LSTM \cite{bamber2025hybrid} & \textbf{96.7}$\pm$0.1 & 94.7$\pm$0.1 & 96.2$\pm$0.2 & 97.7$\pm$0.1 & 96.7$\pm$0.3 & 90.3$\pm$0.3 \\
        & & ResNeXt-50 \cite{ZHAO2026104783} & 96.6$\pm$0.2 & 94.4$\pm$0.3 & 96.4$\pm$0.3 & 97.9$\pm$0.1 & \textbf{97.4}$\pm$0.3 & 91.0$\pm$0.5 \\
    \midrule

    \multirow{9}{*}{\textbf{NSL-KDD}} 
        & \multirow{4}{*}{Original} 
        & CNN \cite{park2022enhanced} & \textbf{79.0}$\pm$1.3 & 78.9$\pm$1.4 & 96.6$\pm$0.6 & 79.8$\pm$1.0 & \textbf{65.6}$\pm$2.7 & 78.0$\pm$1.8 \\
        & & CNN-BiLSTM \cite{jouhari2024lightweight} & 78.0$\pm$0.6 & 78.0$\pm$0.6 & 96.5$\pm$1.9 & 79.1$\pm$0.7 & 64.0$\pm$1.4 & 76.8$\pm$0.7 \\
        & & CNN-LSTM \cite{bamber2025hybrid} & 76.7$\pm$0.2 & 76.6$\pm$0.2 & \textbf{97.8}$\pm$0.1 & 78.3$\pm$0.1 & 60.8$\pm$0.3 & 74.8$\pm$0.2 \\
        & & ResNeXt-50 \cite{ZHAO2026104783} & 79.5$\pm$0.7 & 79.5$\pm$0.8 & 97.4$\pm$1.0 & 80.4$\pm$0.5 & 66.0$\pm$1.7 & 78.6$\pm$1.0 \\
    \cmidrule(lr){2-9}
        & \multirow{4}{*}{Image} 
        & CNN \cite{park2022enhanced} & 79.1$\pm$0.7 & 79.0$\pm$0.7 & \textbf{97.1}$\pm$0.1 & 80.0$\pm$0.6 & 65.3$\pm$1.3 & 78.0$\pm$1.0 \\
        & & CNN-BiLSTM \cite{jouhari2024lightweight} & 78.2$\pm$0.8 & 78.1$\pm$0.9 & 97.0$\pm$0.4 & 79.3$\pm$0.7 & 63.9$\pm$1.7 & 76.9$\pm$1.2 \\
        & & CNN-LSTM \cite{bamber2025hybrid} & \textbf{80.2}$\pm$0.9 & 80.1$\pm$0.9 & \textbf{97.1}$\pm$0.1 & 80.8$\pm$0.8 & \textbf{67.3}$\pm$1.7 & 79.4$\pm$1.2 \\
        & & ResNeXt-50 \cite{ZHAO2026104783} & 78.7$\pm$0.8 & 78.7$\pm$0.8 & \textbf{97.1}$\pm$0.2 & 79.7$\pm$0.6 & 64.8$\pm$1.4 & 77.6$\pm$1.0 \\
    \bottomrule
    \end{tabular*}
    \begin{tablenotes}
        \footnotesize
        \item \textit{Note:} The best accuracy (Acc.) of the IDS models tested with each dataset is highlighted in bold.
    \end{tablenotes}
    \end{threeparttable}
\end{table*}

\begin{table*}[htbp]
    \centering
    \begin{threeparttable}
    \caption{Multi-Classification Results of UNSW-NB15 Dataset With Different IDS Models}
    \label{tab:results_UNSW}
    \scriptsize
    \begin{tabular*}{\textwidth}{@{\extracolsep{\fill}}
        l 
        l 
        l 
        C{0.7cm} 
        C{0.7cm} 
        C{0.8cm} C{0.7cm} 
        C{0.8cm} C{0.8cm} 
        C{0.8cm} C{0.8cm} 
        C{0.7cm} C{0.7cm} 
    }
    \toprule
    \textbf{Dataset} & \textbf{Training Data} & \textbf{IDS Model} 
    & \multicolumn{2}{c}{\textbf{Overall}} 
    & \multicolumn{2}{c}{\textbf{DoS}} 
    & \multicolumn{2}{c}{\textbf{Reconnaissance}} 
    & \multicolumn{2}{c}{\textbf{Shellcode}} 
    & \multicolumn{2}{c}{\textbf{Worms}} \\
    \cmidrule(lr){4-5} \cmidrule(lr){6-7} \cmidrule(lr){8-9} \cmidrule(lr){10-11} \cmidrule(lr){12-13}
    & & 
    & \textbf{Acc.(\%)} & \textbf{F1(\%)} 
    & \textbf{Acc.(\%)} & \textbf{F1(\%)} 
    & \textbf{Acc.(\%)} & \textbf{F1(\%)} 
    & \textbf{Acc.(\%)} & \textbf{F1(\%)} 
    & \textbf{Acc.(\%)} & \textbf{F1(\%)} \\
    \midrule
    
    \multirow{9}{*}{\textbf{UNSW-NB15}}
        & \multirow{4}{*}{Original}
        & CNN \cite{park2022enhanced} & \textbf{86.5}$\pm$0.8 & 51.5$\pm$1.8  & \textbf{64.6}$\pm$9.4 & 68.5$\pm$6.0 & 43.7$\pm$3.0 & 48.1$\pm$1.9 & \textbf{45.8}$\pm$4.0   & 24.5$\pm$1.3   & 21.1$\pm$4.4   & 23.6$\pm$2.6   \\
        & & CNN-BiLSTM \cite{jouhari2024lightweight} & 81.1$\pm$0.5 & 36.1$\pm$1.7 & 19.6$\pm$1.4 & 26.2$\pm$1.5 & 33.0$\pm$1.8 & 39.7$\pm$1.9 & 18.7$\pm$5.3 & 16.8$\pm$1.6 & 4.8$\pm$4.6 & 8.5$\pm$7.8 \\
        & & CNN-LSTM \cite{bamber2025hybrid} & 82.6$\pm$0.5 & 44.8$\pm$1.1   & 45.7$\pm$1.8 & 48.2$\pm$1.7 & 30.2$\pm$0.3 & 40.1$\pm$0.8 & 32.7$\pm$0.7 & 19.9$\pm$0.9 & 16.4$\pm$4.3 & 25.6$\pm$5.7 \\
        & & ResNeXt-50 \cite{ZHAO2026104783} & 84.6$\pm$1.0 & 49.6$\pm$1.7  & 58.9$\pm$14.3 & 57.7$\pm$10.1 & \textbf{43.9}$\pm$5.9 & 50.3$\pm$5.7 & 31.2$\pm$6.9 & 20.0$\pm$3.3 & \textbf{22.7}$\pm$4.0 & 28.9$\pm$7.0 \\
    \cmidrule(lr){2-13} 
        & \multirow{4}{*}{Image}
        & CNN \cite{park2022enhanced} & 92.8$\pm$3.5 & 67.7$\pm$16.6  & 81.0$\pm$28.5 & 73.3$\pm$25.8 & 75.5$\pm$26.5  & 79.0$\pm$27.8  & 77.5$\pm$27.3 & 34.9$\pm$12.3 & \textbf{62.3}$\pm$22.1 & 54.6$\pm$19.4 \\
        & & CNN-BiLSTM \cite{jouhari2024lightweight} & 93.9$\pm$0.3 & 73.7$\pm$1.2 & 69.7$\pm$8.7 & 70.0$\pm$5.3 & 52.8$\pm$4.3 & 53.1$\pm$3.1 & 36.9$\pm$4.3  & 23.7$\pm$3.5  & 13.0$\pm$7.0 & 13.5$\pm$5.7 \\
        & & CNN-LSTM \cite{bamber2025hybrid} & 93.7$\pm$0.2 & 72.5$\pm$1.0   & 73.8$\pm$1.2 & 69.8$\pm$1.8 & 35.1$\pm$2.6 & 44.0$\pm$2.9 & 36.9$\pm$3.2 & 21.3$\pm$1.3 & 20.0$\pm$5.3 & 30.9$\pm$6.0 \\
        & & ResNeXt-50 \cite{ZHAO2026104783} & \textbf{95.3}$\pm$0.3 & 75.9$\pm$1.8   & \textbf{92.0}$\pm$1.0 & 84.8$\pm$0.5 & \textbf{84.6}$\pm$0.5 & 89.9$\pm$0.4 & \textbf{82.6}$\pm$3.6 & 44.4$\pm$4.6 & 61.6$\pm$9.3 & 62.2$\pm$6.6 \\
    \bottomrule
    \end{tabular*}
    \begin{tablenotes}
        \footnotesize
        \item \textit{Note:} The best accuracy (Acc.) of the IDS models is highlighted in bold.
    \end{tablenotes}
    \end{threeparttable}
\end{table*}

\begin{table*}[htbp]
    \centering
    \begin{threeparttable}
    \caption{Multi-Classification Results of NSL-KDD Dataset With Different IDS Models}
    \label{tab:NSL_results}
    \scriptsize
    \begin{tabular*}{\textwidth}{@{\extracolsep{\fill}}
        l 
        l 
        l 
        C{0.7cm} 
        C{0.9cm} 
        C{0.6cm} C{1.0cm} 
        C{0.6cm} C{1.0cm} 
        C{0.6cm} C{1.0cm} 
        C{0.6cm} C{1.0cm} 
    }
    \toprule
    \textbf{Dataset} & \textbf{Training Data} & \textbf{IDS Model} 
    & \multicolumn{2}{c}{\textbf{Overall}} 
    & \multicolumn{2}{c}{\textbf{R2L}} 
    & \multicolumn{2}{c}{\textbf{Probe}} 
    & \multicolumn{2}{c}{\textbf{DoS}} 
    & \multicolumn{2}{c}{\textbf{U2R}} \\
    \cmidrule(lr){4-5} \cmidrule(lr){6-7} \cmidrule(lr){8-9} \cmidrule(lr){10-11} \cmidrule(lr){12-13}
    & & 
    & \textbf{Acc.(\%)} & \textbf{F1(\%)} 
    & \textbf{Acc.(\%)} & \textbf{F1(\%)} 
    & \textbf{Acc.(\%)} & \textbf{F1(\%)} 
    & \textbf{Acc.(\%)} & \textbf{F1(\%)} 
    & \textbf{Acc.(\%)} & \textbf{F1(\%)} \\
    \midrule
    
    \multirow{9}{*}{\textbf{NSL-KDD}}
        & \multirow{4}{*}{Original}
        & CNN \cite{park2022enhanced} & 76.8$\pm$0.3 & 56.6$\pm$1.4  & 1.8$\pm$0.6 & 3.5$\pm$1.1 & \textbf{77.7}$\pm$2.8 & 78.5$\pm$1.7 & 80.0$\pm$1.5   & 87.6$\pm$0.9   & 21.1$\pm$4.2   & 33.8$\pm$5.6   \\
        & & CNN-BiLSTM \cite{jouhari2024lightweight} & 75.1$\pm$1.5 & 53.5$\pm$2.5 & 4.2$\pm$2.3 & 7.9$\pm$4.2 & 58.8$\pm$4.5 & 65.4$\pm$4.5 & 79.1$\pm$3.4 & 86.8$\pm$2.1 & 18.4$\pm$7.0 & 28.6$\pm$8.4 \\
        & & CNN-LSTM \cite{bamber2025hybrid} & \textbf{77.0}$\pm$0.6 & 55.3$\pm$1.0   & \textbf{6.2}$\pm$1.2 & 11.6$\pm$2.1 & 54.4$\pm$4.1 & 65.9$\pm$3.6 & \textbf{85.6}$\pm$0.2 & 90.6$\pm$0.2 & 17.9$\pm$3.5 & 29.1$\pm$4.9 \\
        & & ResNeXt-50 \cite{ZHAO2026104783} & 76.5$\pm$1.1 & 57.5$\pm$2.8   & 10.1$\pm$7.1 & 17.2$\pm$10.4 & 56.8$\pm$3.8 & 66.7$\pm$3.1 & 81.6$\pm$1.2 & 87.9$\pm$1.1 & \textbf{25.8}$\pm$6.9 & 36.1$\pm$6.9 \\
    \cmidrule(lr){2-13} 
        & \multirow{4}{*}{Image}
        & CNN \cite{park2022enhanced} & 78.5$\pm$0.7 & 65.8$\pm$1.9  & 35.0$\pm$5.2 & 51.3$\pm$5.8 & 65.3$\pm$1.8  & 70.7$\pm$1.4  & 76.5$\pm$0.5 & 85.3$\pm$0.3 & 36.1$\pm$5.9 & 40.9$\pm$4.9 \\
        & & CNN-BiLSTM \cite{jouhari2024lightweight} & 77.8$\pm$0.7 & 64.4$\pm$2.0 & 27.7$\pm$3.5 & 42.9$\pm$4.2 & 65.9$\pm$1.8 & 71.5$\pm$1.2 & 76.5$\pm$1.9  & 85.0$\pm$1.0  & 36.1$\pm$7.1 & 41.9$\pm$8.5 \\
        & & CNN-LSTM \cite{bamber2025hybrid} & \textbf{80.2}$\pm$0.6 & 68.1$\pm$1.4   & \textbf{41.8}$\pm$4.3 & 58.3$\pm$4.1 & \textbf{69.8}$\pm$2.0 & 73.5$\pm$1.4 & \textbf{77.0}$\pm$0.4 & 85.6$\pm$0.3 & \textbf{37.0}$\pm$5.9 & 40.7$\pm$8.4 \\
        & & ResNeXt-50 \cite{ZHAO2026104783} & 77.3$\pm$1.3 & 60.7$\pm$2.8   & 28.9$\pm$7.3 & 44.3$\pm$8.8 & 62.1$\pm$3.9 & 69.6$\pm$2.2 & 75.3$\pm$3.1 & 84.5$\pm$1.8 & 34.1$\pm$9.5 & 24.9$\pm$13.8 \\
    \bottomrule
    \end{tabular*}
    \begin{tablenotes}
        \footnotesize
        \item \textit{Note:} The best accuracy (Acc.) of the IDS models is highlighted in bold.
    \end{tablenotes}
    \end{threeparttable}
\end{table*}

Table~\ref{tab2} presents the binary classification results of the four IDS models on UNSW-NB15 and NSL-KDD.

On UNSW-NB15, the IDS models trained on image-encoded inputs consistently outperform their flow-based counterparts by a substantial margin, with overall accuracy improvement by 14.6\% to 15.7\% across all four models, and the CNN-LSTM-based IDS model achieving the highest accuracy of 96.7\%. The IDS models trained on flow-based inputs exhibit a pronounced performance disparity between classes: normal class accuracy ranges from 89.1\% to 91.9\%, while attack class accuracy is considerably lower, ranging from 29.2\% to 48.1\%, indicating that 1D models are biased towards the majority class. The proposed image encoding substantially alleviates this disparity: attack class accuracy rises above 96.7\% for all IDS models, with the CNN-BiLSTM-based IDS model reaching 97.1\%, and attack F1-scores improve from 35.0\%--48.7\% to 90.3\%--91.4\%, an absolute improvement of over 40\%. Standard deviations are also notably smaller under the image modality, suggesting improved training stability.

On NSL-KDD, performance gains are more varied across the models. The CNN-LSTM-based IDS model benefits the most, with accuracy increasing from 76.7\% to 80.2\%, attack class accuracy from 60.8\% to 67.3\%, and attack F1-score from 74.8\% to 79.4\%. The CNN- and CNN-BiLSTM-based IDS models show marginal improvements of 0.1\% and 0.2\%, respectively. The ResNeXt-50-based IDS model exhibits a slight decrease of 0.8\%, from 79.5\% to 78.7\%, attributable to its flow-based counterpart already achieving the highest normal class accuracy (97.4\%) and attack class accuracy (66.0\%) among 1D models, leaving limited room for further improvement. The smaller gains on NSL-KDD reflect the dataset's higher proportion of binary and nominal features, which occupy fewer bytes and produce more zero-padded regions, reducing the density of meaningful spatial information available to convolutional filters.

\subsection{Multi-classification}

Tables~\ref{tab:results_UNSW} and~\ref{tab:NSL_results} present the multi-classification results on UNSW-NB15 and NSL-KDD, respectively.

On UNSW-NB15, the IDS models trained on image-encoded inputs achieve accuracy improvements of 6.3\% to 12.8\% over their flow-based counterparts. The ResNeXt-50-based IDS model achieves the highest accuracy of 95.3\%, with per-class accuracy of 92.0\% for DoS, 84.8\% for Reconnaissance, 82.6\% for Shellcode, and 61.6\% for Worms. Here, a particularly remarkable improvement over flow-based results has been noted, where Shellcode and Worms accuracy remained below 31.2\% and 22.7\%, respectively. The CNN-BiLSTM- and CNN-LSTM-based IDS models show comparable overall gains (12.8\% and 11.1\%), with strong improvements on DoS and Reconnaissance but limited gains on Shellcode and Worms. This suggests that lighter architectures may have insufficient capacity to exploit spatial patterns for the most underrepresented classes. The CNN-based IDS model achieves 92.8\% accuracy but with high standard deviations, which indicates instability that is likely attributable to its shallower architecture.

On NSL-KDD, overall accuracy improvements range from 0.8\% to 3.2\%, with the CNN-LSTM-based IDS model again showing the largest gain (77.0\% to 80.2\%). The most notable finding is the substantial improvement in minority class detection. R2L accuracy increases from 1.8\% to 35.0\% for the CNN-based IDS model and from 6.2\% to 41.8\% for the CNN-LSTM-based IDS model; U2R accuracy improves from 21.1\% to 36.1\% and from 17.9\% to 37.0\%, respectively. These gains are reflected in average F1-scores, representing absolute improvements of 9.2\% and 12.8\% for the CNN- and CNN-LSTM-based IDS models, respectively. However, DoS accuracy decreases slightly with the image modality for all models, suggesting the encoding redistributes discriminative capacity towards minority classes at the expense of the majority attack class.

Across both datasets, the image encoding provides the greatest benefit for classes that are poorly detected with the flow modality. On UNSW-NB15, where 1D models struggle with all attack classes, improvements are broad and substantial. On NSL-KDD, where DoS and Probe are already reasonably well detected, the encoding primarily benefits R2L and U2R. This pattern suggests that byte-level spatial representation enables 2D convolutional filters to capture inter-feature relationships inaccessible through sequential 1D processing, with the greatest advantage when 1D representations lack sufficient discriminative information. Although some of the prior research reports higher absolute accuracy on the same datasets, direct comparison is not straightforward, as those results were obtained under dataset-specific conditions that differ substantially from the unified evaluation settings adopted here.

\subsection{Limitations}

While the proposed encoding demonstrates consistent improvements across both datasets, a few limitations remain. First, its effectiveness is partially dependent on feature composition: NSL-KDD contains a higher proportion of binary and nominal features, which occupy fewer bytes and produce more zero-padded regions, reducing the density of meaningful spatial information available to convolutional filters. Second, the fixed sequential layout does not account for inter-feature correlations, meaning semantically related features may occupy spatially distant image regions, potentially limiting the ability of local convolutional filters to exploit their relationships. Third, all experiments are conducted on offline benchmark datasets, and performance in real-time or streaming traffic conditions remains to be validated.

\section{Conclusion}

This paper presents a novel byte-level flow-to-image encoding method for deep learning-based IDS models, converting tabular network records into fixed-size 32×32 RGB images. The encoding is deterministic and reversible, preserving a fixed spatial layout that enables 2D convolutional filters to capture inter-feature relationships not accessible through sequential 1D processing. Experimental evaluations on NSL-KDD and UNSW-NB15 demonstrate consistent improvements across four IDS models, with accuracy gains of up to 15.6\% and 12.8\% for binary and multi-classification on UNSW-NB15, and up to 3.5\% and 3.2\% on NSL-KDD. These results demonstrate that the proposed encoding strengthens the ability of IDS models to detect both common and rare attack types in local computer networks, where timely and accurate intrusion detection is critical for network reliability and security.

In future work, we aim to explore adaptive spatial layouts that optimise feature placement based on inter-feature correlation, including a principled investigation of the relative positioning of discrete and continuous feature regions, which in the current design are determined empirically. We also plan to extend the evaluation to real-world network traffic traces and investigate deployment of the encoding pipeline in lightweight, resource-constrained network monitoring systems.

\bibliographystyle{IEEEtran} 
\bibliography{references}

\end{document}